\documentclass[preprint,showpacs,preprintnumbers,amsmath,amssymb]{revtex4}

\usepackage{graphicx}
\usepackage{dcolumn}
\usepackage{bm}

\begin{document}
\title{The N=28 shell closure; from N=Z to the neutron drip line}
\author{E. Caurier}
\email{etienne.caurier@ires.in2p3.fr}
\author{F. Nowacki}
\email{frederic.nowacki@ires.in2p3.fr}
\address{Groupe de Physique Th\'eorique, IReS, B\^at. 27, 
    IN2P3-CNRS/Universit\'e Louis Pasteur, BP 28,
  F--67037 Strasbourg Cedex~2, France}
\author{A. Poves}
\email{alfredo.poves@uam.es}
\address{Departamento de F\'{\i}sica Te\'orica C-XI, Universidad
  Aut\'onoma de Madrid, E--28049 Madrid, Spain}


\begin{abstract}
The evolution of the N=28 neutron shell closure is studied in the N=28 isotones
from neutron drip line's $^{40}$Mg to  N=Z doubly magic $^{56}$Ni. It is found that
the N=28 closure vanishes at Z=12 in favor of a deformed ground state. For Z=14 and Z=16
closed shell and intruder configurations are almost degenerate,
leading to  
highly  mixed ground states  and to very low lying excited 0$^+$ states. For Z$>$20 the intruder 
states are always above the closed shell ground states. We examine their
structure  and their  possible existence  as collective yrare bands of well defined particle-hole character.   
\end{abstract}


\maketitle

\section{Introduction}
\label{sec:intro}

The fate of the magic closures far from stability is an issue of great current 
experimental and theoretical activity. Harmonic oscillator shell closures have been 
shown not to hold in the neutron rich regions ($^{11}$Li and $^{12}$Be at N=8;  
$^{31}$Na and $^{32}$Mg at N=20; $^{68}$Ni at N=40) as well as in the proton rich 
side ($^{80}$Zr at N=Z=40). Spin-orbit closures (28, 50 82, 126) 
seemed until now to be more robust. How can a shell closure disappear? It depends on 
the balance of two opposite tendencies. On one side magic numbers are associated to
energy gaps in the spherical mean field, this means that to promote particles  
above a  closed shell, costs energy. However, this energy can be partly recovered,
 because closed shells have no correlation energy, while open shell configurations
of neutrons and protons have a lot. For instance, in  
$^{80}$Zr, the gap between the $pf$-shell and the 1g$_{9/2}$ orbit is greatly reduced as a
 consequence of the very attractive spin-orbit interaction. This makes it energetically
 favourable to promote particles beyond the $pf$-shell to take advantage of the large \
quadrupole and pairing correlations. Indeed, $^{80}$Zr is found experimentally to be
a well deformed rotor, consistent with a large occupancy of the 1g$_{9/2}$ orbit and
 its quadrupole partners.   
 
 At the neutron rich edge, the structure of the spherical mean field may be at variance 
 with the usual one at the stability line. The reason is that at the stability line the T=0
 channel of the nucleon-nucleon interaction has a stronger weight relative to the T=1
 channel than it has when the neutron excess is very large. When some of the gaps get 
 reduced, open shell configurations, usually two neutron excitations across the neutron
 closure, take advantage of the availability of open shell protons to build highly
  correlated states that are more bound than the closed shell configuration. 
 Then the shell closure is said to have vanished. $^{31}$Na is a reference case for this 
physics.
 
  In this article we shall address the question of the persistence of the N=28 shell 
  closure. It is the first spin-orbit driven ``classical'' closure. In combination with the N=20
  harmonic oscillator closure it produces the doubly magic $^{48}$Ca, while at N=Z leads 
  to    $^{56}$Ni.  $^{68}$Ni, another potential doubly magic nuclei does not appear to
  be one, while  $^{78}$Ni is believed to be doubly magic again.
   In addition we shall examine the structure of the states that can
  compete with, and even beat 
  the magic closure, whether they correspond to intrinsic structures or not and whether 
  they keep their identity in the spectra even if they do not become ground states.

\section{The N=28 isotones}
\label{sec:n28}

  Our starting point is $^{48}$Ca, one of the strongest closures in the nuclear chart.
  To the proton rich side we can reach  $^{56}$Ni, while to the neutron rich side we could 
  aim to doubly magic  $^{36}$O, were it not because this nucleus is well beyond the
  neutron drip line, as it is also the case of $^{38}$Ne. For Z$\le$20  we use a valence 
  space comprising the $pf$-shell for the neutrons and the $sd$-shell for the protons. The
  interaction is described in \cite{Nummela.ea:2001}. For Z$>$20 we use the $pf$-shell 
  for neutrons and protons and the interaction KB3G \cite{A50}.    
  The intruder configurations are characterised  by imposing that two  neutrons be 
  {\it outside} the 1f$_{7/2}$ neutron orbit.

 The diagonalizations are performed in the $m$-scheme using a fast
 implementation of the Lanczos algorithm through the code {\sc
 antoine}~\cite{antoine,nathan} or in  J-coupled scheme using the code
 {\sc nathan} \cite{nathan}. Some details may be found in
 ref.~\cite{masses}.  The Lanczos Strength Function (LSF) method,
 following  Whitehead's prescription~\cite{white}, is explained and 
 illustrated in refs.~\cite{cpz1,cpz2,bloom}. 

 We shall examine first the results gathered in Table~\ref{tab:n28gap}. In our reference 
 magic nucleus, $^{48}$Ca, 
 the intruder configuration is 5.38~MeV less bound than the 
 closed shell. This number results of subtracting two contributions:
 Promoting two particles across the N=28 neutron closure, represents an energy loss of
 twice the value of the gap (4.73~MeV in our calculation). On the other side, the intruder
 configuration gains 4.08~MeV of correlation energy (mainly due to pairing) relative to
 the closed shell. When full mixing is allowed, the ground state of  
 $^{48}$Ca remains a closed shell at 90\%.

\begin{table}
\caption{N=28 isotones: quasiparticle neutron gaps, difference in
  correlation energies between the 2p-2h and the 0p-0h configurations
  and their relative position}
\label{tab:n28gap}       
 \begin{tabular*}{\linewidth}{@{\extracolsep{\fill}}lccccccccc} 
\hline\noalign{\smallskip}
& $^{40}$Mg 
& $^{42}$Si 
& $^{44}$S
& $^{46}$Ar 
& $^{48}$Ca
& $^{50}$Ti
& $^{52}$Cr
& $^{54}$Fe 
& $^{56}$Ni \\
\noalign{\smallskip}\hline\noalign{\smallskip}
gap         & 3.35    & 3.50     & 3.23  &  3.84  & 4.73 & 5.33 &
5.92 & 6.40 & 7.12 \\
$\Delta$E$_{\mathcal{C}orr}$   & 8.45    & 6.0      & 6.66  &  5.98  &
4.08 & 7.59 & 10.34 & 10.41 & 6.19 \\
E$^*_{2p-2h}$ & -1.75   & 1.0      & -0.2  &  1.7   & 5.38 &
3.07 & 1.50 & 2.39 & 8.05 \\
\noalign{\smallskip}\hline
 \end{tabular*}
\end{table}

Adding protons to $^{48}$Ca increases the neutron gap. The correlation
energy of the intruder state grows also rapidly to reach 10.32~MeV in
$^{52}$Cr, it remains almost constant in $^{54}$Fe and drops abruptly
at the proton closure in $^{56}$Ni. In all the cases the intruder
configuration is less bound than the closed shell. It is at mid-proton
shell where the energy difference is smaller (1.5~MeV). After mixing,
the first excited 0$^+$ state of $^{52}$Cr, dominated by the intruder
configurations, lies at 2.43~MeV, compared to the experimental value
2.65~MeV. Similar degree of accord with experiment is found in
$^{54}$Fe (2.80~MeV $vs$ 2.56~MeV) and $^{50}$Ti (4.08~MeV $vs$
3.87~MeV). The case of $^{56}$Ni is different because the first
excited 0$^+$ state is predominantly 4p-4h. Its correlation energy is
very large. Depending on the effective interaction used, the fully
mixed results give a 60-70\% of doubly magic closure its ground state.

When protons are removed from $^{48}$Ca, then N=28 neutron gap first
diminishes and then remains constant at about 3.5~MeV. The correlation
energy of the intruder increases in $^{46}$Ar and $^{44}$S and then
goes down slightly at the proton sub-shell closure in $^{42}$Si.
Finally it rises abruptly at $^{40}$Mg. The resulting scenarios are 
quite diverse. In $^{46}$Ar the intruder is clearly above the closed
shell, in $^{44}$S they are almost degenerate, in $^{42}$Si the closed
shell takes over again by little and in $^{40}$Mg the intruder state
is well below the closed shell.

\section{The structure of the intruder states}
\label{sec:intru}

The states that challenge the shell closure have large correlation
energy. Large correlation energy is most often related to large
quadrupole coherence. If this were the case here, the vanishing of the
N=28 shell closure would be accompanied of a transition from spherical
to deformed shapes or would result in a strong case of coexistence.

In order to understand this issue we first calculate the spin
sequences 0$^+$, 2$^+$, 4$^+$, etc.  in the basis of all the
configurations that have two neutrons outside the 1f$_{7/2}$ orbit.
Our conclusions about the collective character of the bands will be
based on the analysis of the excitation energies, the static
quadrupole moments and the E2 transition probabilities, that we have
gathered in Table~\ref{tab:n28ph1} for Z$>$20 and in
Table~\ref{tab:n28ph2} for Z$<$20.
  
  In $^{54}$Fe we face a clear-cut example of rotational structure
  that stands up to spin J=8.  The static and dynamic (transition)
  quadrupole moments, which can be extracted from the spectroscopic
  quadrupole moments and B(E2)'s using the standard formulae (see
  ref.~\cite{Bohr.Mottelson:1975}), are nicely constant and correspond
  roughly to a deformation $\beta$=0.3. We can therefore conclude that
  the lowest intruder band in $^{54}$Fe has a prolate intrinsic
  structure. How it manifests in the physical spectrum will be the
  topic of the next section. Similar conclusions hold when we move to
  $^{52}$Cr, whose intruder state is an even better rotor and slightly
  more deformed.  On the contrary, the intruder band in $^{50}$Ti is
  less convincing. It has some rotational flavor, but not enough to
  conclude that it represent a ``bona fide'' intrinsic state.

\begin{table}[h]
\begin{center}
\caption{Properties of the intruder states of 2p-2h character in
  $^{54}$Fe, $^{52}$Cr and $^{50}$Ti. Excitation  energies in MeV,  B(E2)'s in $e^2\,fm^4$,
  Spectroscopic quadrupole moments in  $e\,fm^2$.}
\label{tab:n28ph1}
\vspace{0.3cm}
\begin{tabular*}{\linewidth}{@{\extracolsep{\fill}}cccccccccc}
\hline\hline
     &\multicolumn{3}{c}{$^{54}$Fe} 
     &\multicolumn{3}{c}{$^{52}$Cr} &\multicolumn{3}{c}{$^{50}$Ti} \\
\cline{2-4}\cline{5-7}\cline{8-10}
J & $\Delta$E & $Q$ & B(E2)$\downarrow$ 
  & $\Delta$E & $Q$ & B(E2)$\downarrow$ & $\Delta$E & $Q$  & B(E2)$\downarrow$ \\
\hline 
0 & 0.00 &       &      & 0.00 &       &     & 0.00 &       & \\
2 & 0.51 & -30.0 & 224  & 0.40 & -30.7 & 235 & 0.53 & -20.6 & 128   \\
4 & 1.51 & -36.9 & 312  & 1.19 & -38.5 & 331 & 1.39 & -23.8 & 188  \\
6 & 2.85 & -38.3 & 321  & 2.33 & -40.0 & 339 & 2.36 & -22.6 & 185  \\
8 & 4.48 & -35.0 & 294  & 3.81 & -37.3 & 319 & 3.51 & -26.6 & 161  \\
\hline\hline  
\end{tabular*}
\end{center}  
\end{table}

\begin{table}[h]
\begin{center}
\caption{Properties of the intruder states of 2p-2h character in
  $^{46}$Ar, $^{44}$S, $^{42}$Si and $^{40}$Mg. Excitation  energies in MeV,
  B(E2)'s in $e^2\,fm^4$,
  Spectroscopic quadrupole moments in  $e\,fm^2$.}
\label{tab:n28ph2}
\vspace{0.3cm}
\begin{tabular*}{\linewidth}{@{\extracolsep{\fill}}ccccccccccccc}
\hline\hline
     &\multicolumn{3}{c}{$^{46}$Ar} &\multicolumn{3}{c}{$^{44}$S } 
     &\multicolumn{3}{c}{$^{42}$Si} &\multicolumn{3}{c}{$^{40}$Mg} \\
\cline{2-4}\cline{5-7}\cline{8-10}\cline{11-13}
J & $\Delta$E & $Q$ & B(E2)$\downarrow$ & $\Delta$E & $Q$  & B(E2)$\downarrow$ 
  & $\Delta$E & $Q$ & B(E2)$\downarrow$ & $\Delta$E & $Q$  & B(E2)$\downarrow$  \\
\hline
0 & 0.00 &       &    & 0.00 &       &      & 0.00 &       &    & 0.00 &       &    \\
2 & 0.81 & -8.70 & 61 & 0.73 & -19.0 & 97   & 0.96 & -9.44 & 38 & 0.65 & -19.3 & 95  \\
4 & 2.00 &  0.82 & 80 & 2.13 & -21.5 & 129  & 2.13 & -4.60 & 38 & 1.92 & -22.2 & 125 \\
6 & 2.75 &  22.5 & 41 & 4.16 & -10.0 & 100  & 2.62 & 18.3  & 15 & 3.44 & -0.89 & 78  \\
8 & 4.45 &  21.2 & 81 & 6.73 & -7.74 & 64   & 4.70 &  9.18 & 21 & 5.01 & -4.48 & 74  \\
\hline\hline  
\end{tabular*}
\end{center}  
\end{table}

The situation becomes more involved in the Z$<$20 isotones, because
of the influence of the filling of the proton orbits on the structure
of the intruders. In $^{46}$Ar and $^{42}$Si there is no signal of any
intrinsic state. On the contrary, in $^{44}$S and $^{40}$Mg the
sequence 0$^+$, 2$^+$, 4$^+$ exhibits clear rotational features
corresponding to $\beta$=0.35.  The rotational behaviour of the bands
is somehow blurred at J=6-8. Indeed, for the lower spins we are
entitled to speak of deformed intruder bands.

\begin{figure}[htb]
\begin{center}
  \includegraphics[width=0.6\linewidth]{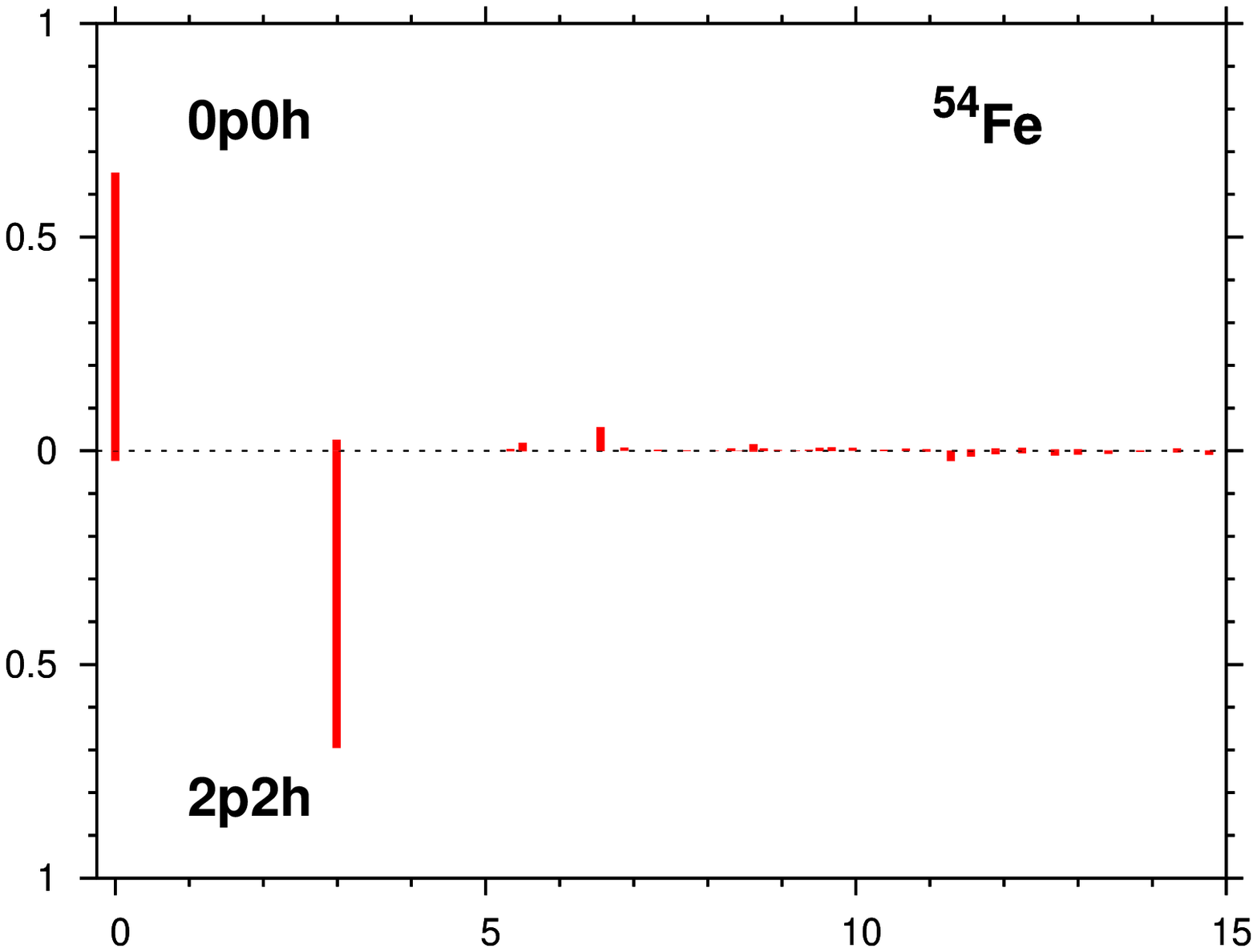}
  \includegraphics[width=0.6\linewidth]{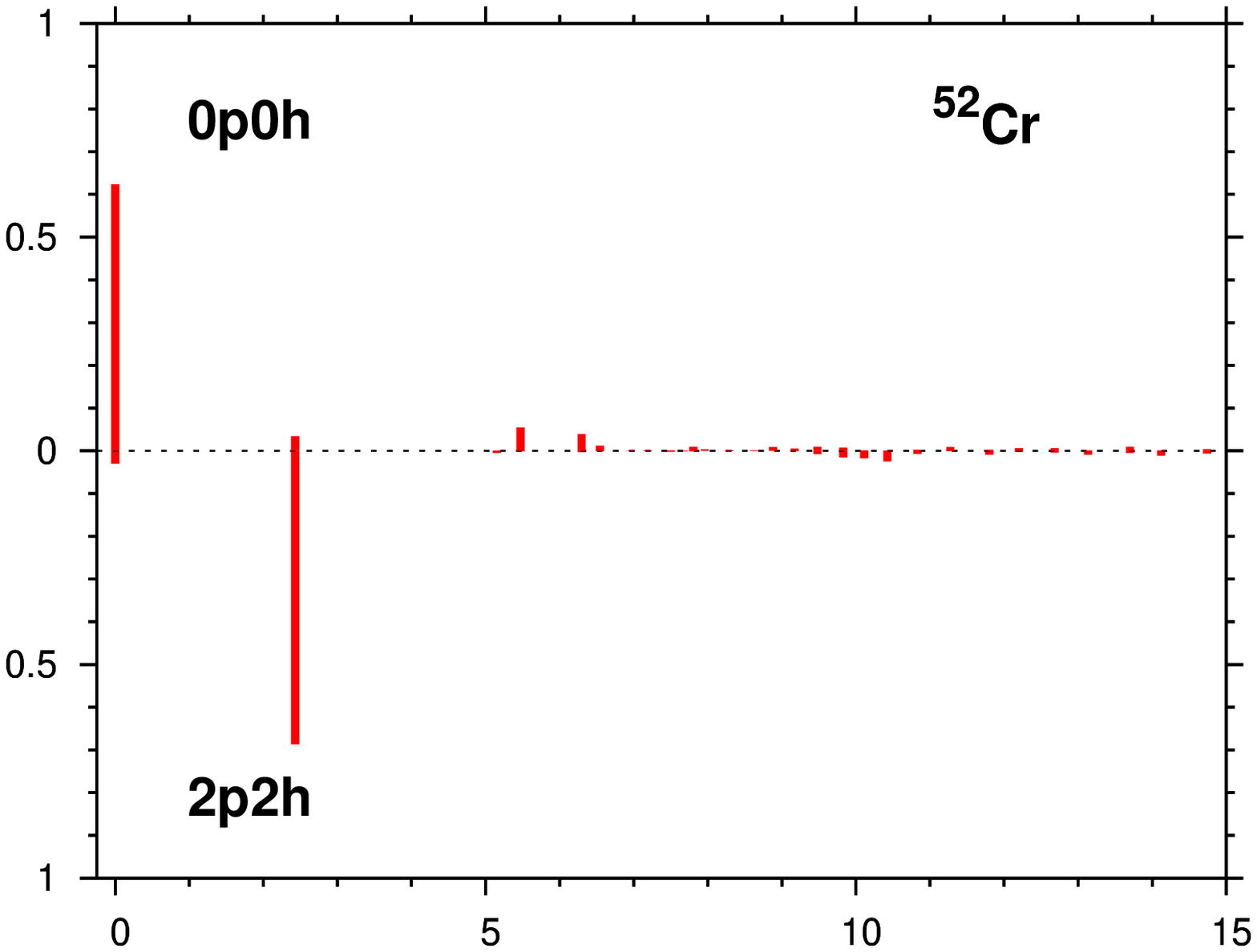}
  \includegraphics[width=0.6\linewidth]{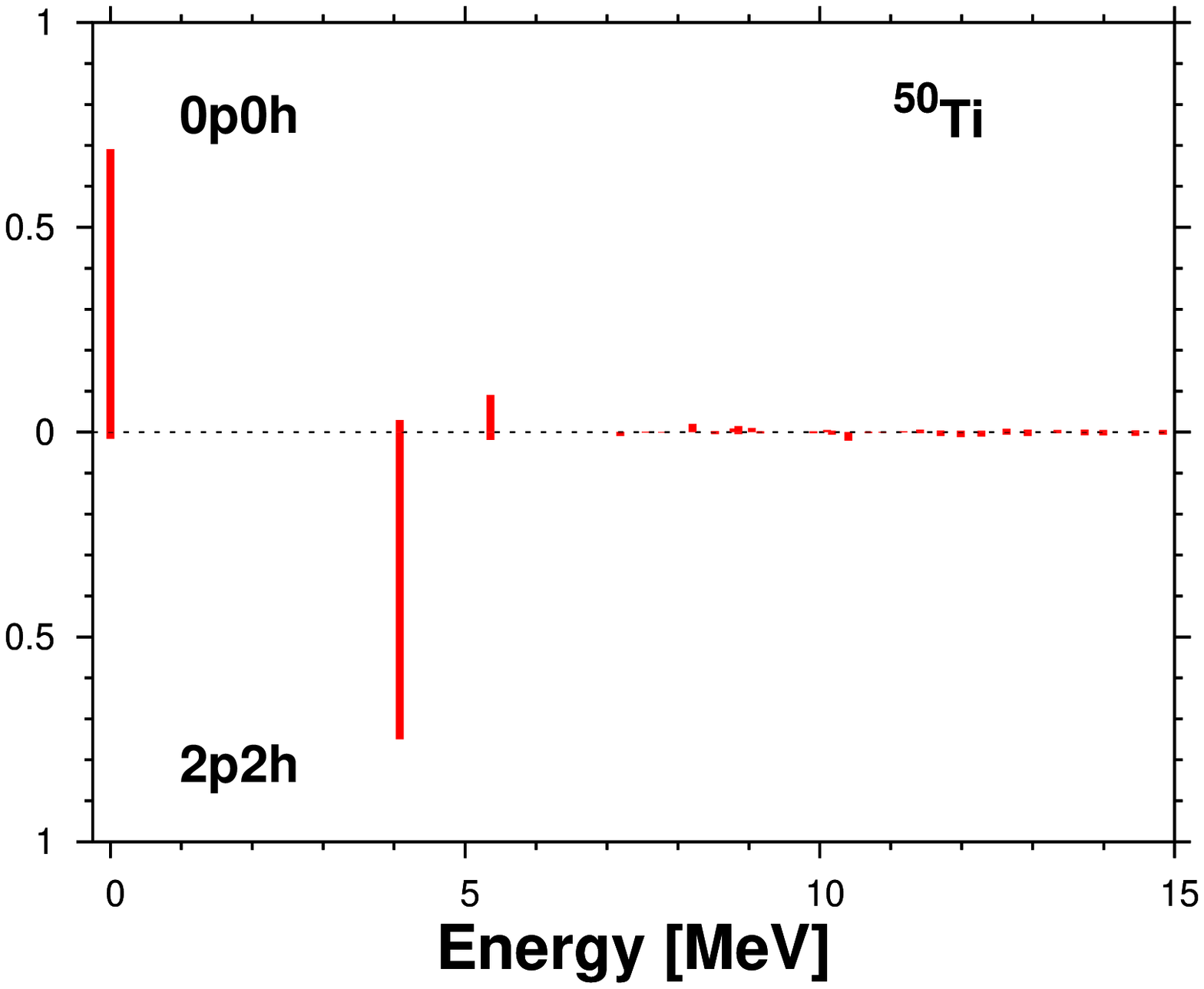}
    \caption{Overlaps (squared) of the closed shell 0$^+$ (upper panels) and the intruder 
     0$^+$ (bottom panels) with the physical 0$^+$ states in $^{50}$Ti, $^{52}$Cr and $^{54}$Fe}
    \label{fig:over1}
\end{center}
\end{figure}

\begin{figure}[htb]
\begin{center}
  \includegraphics[width=0.5\linewidth]{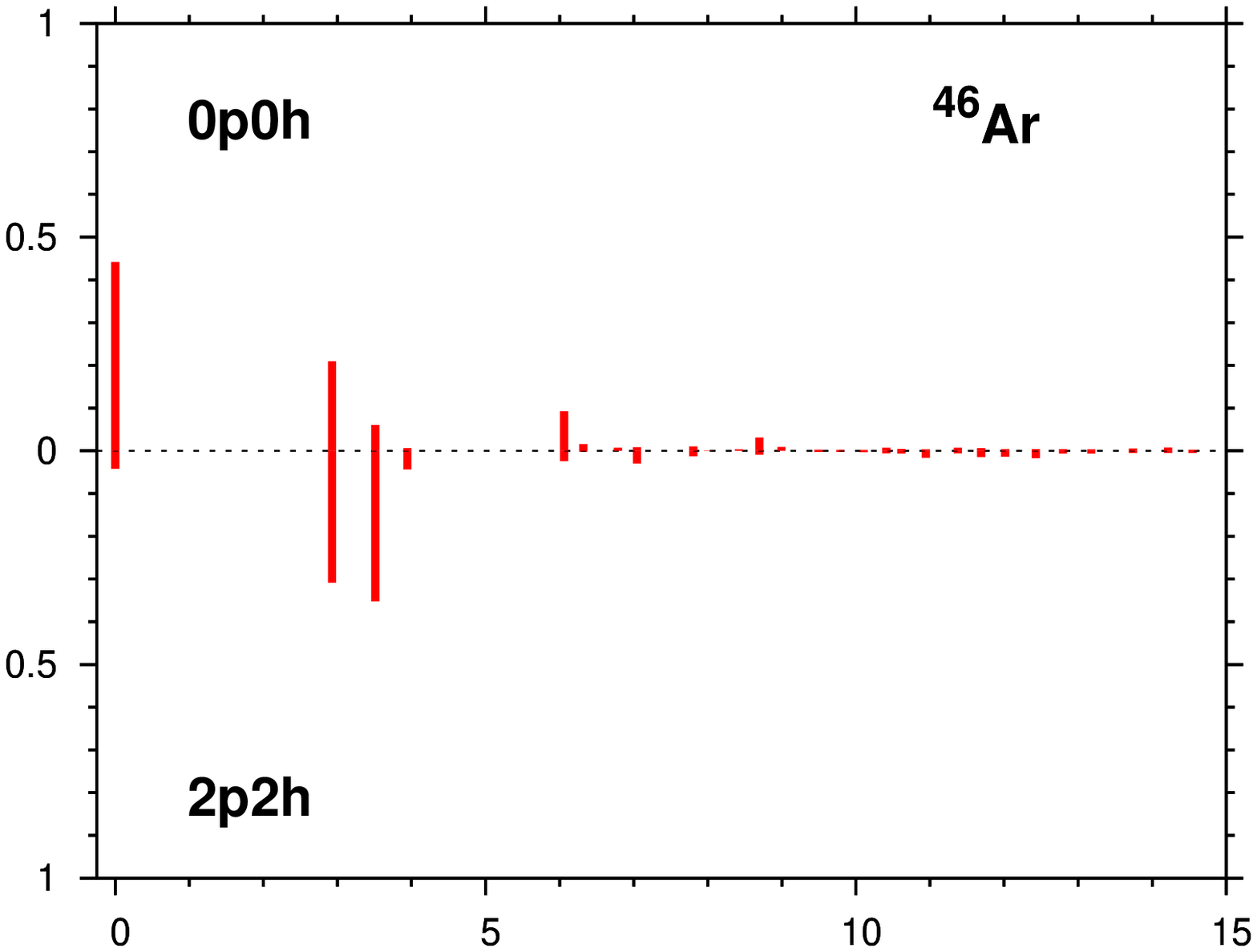}
  \includegraphics[width=0.5\linewidth]{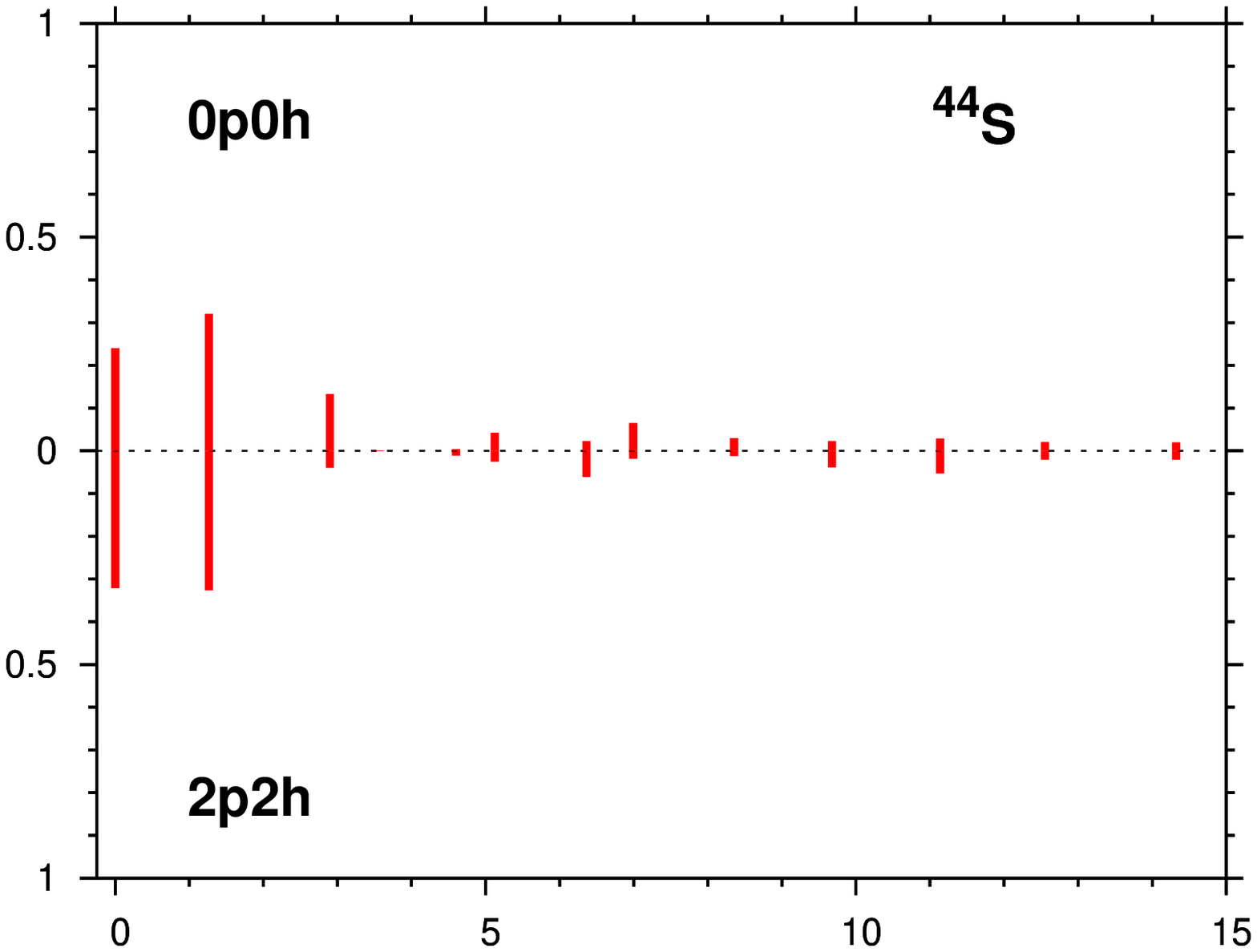}
  \includegraphics[width=0.5\linewidth]{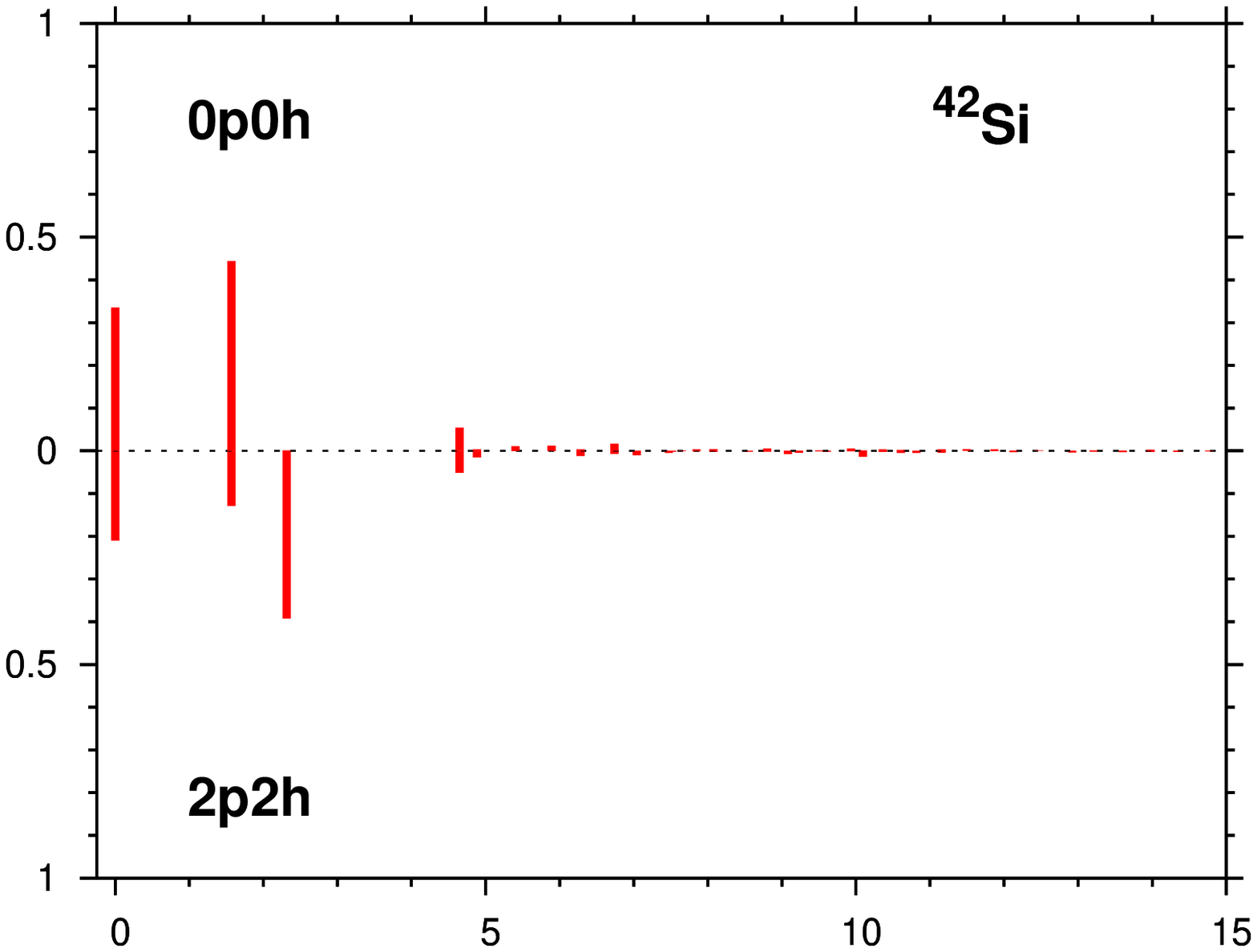}
  \includegraphics[width=0.5\linewidth]{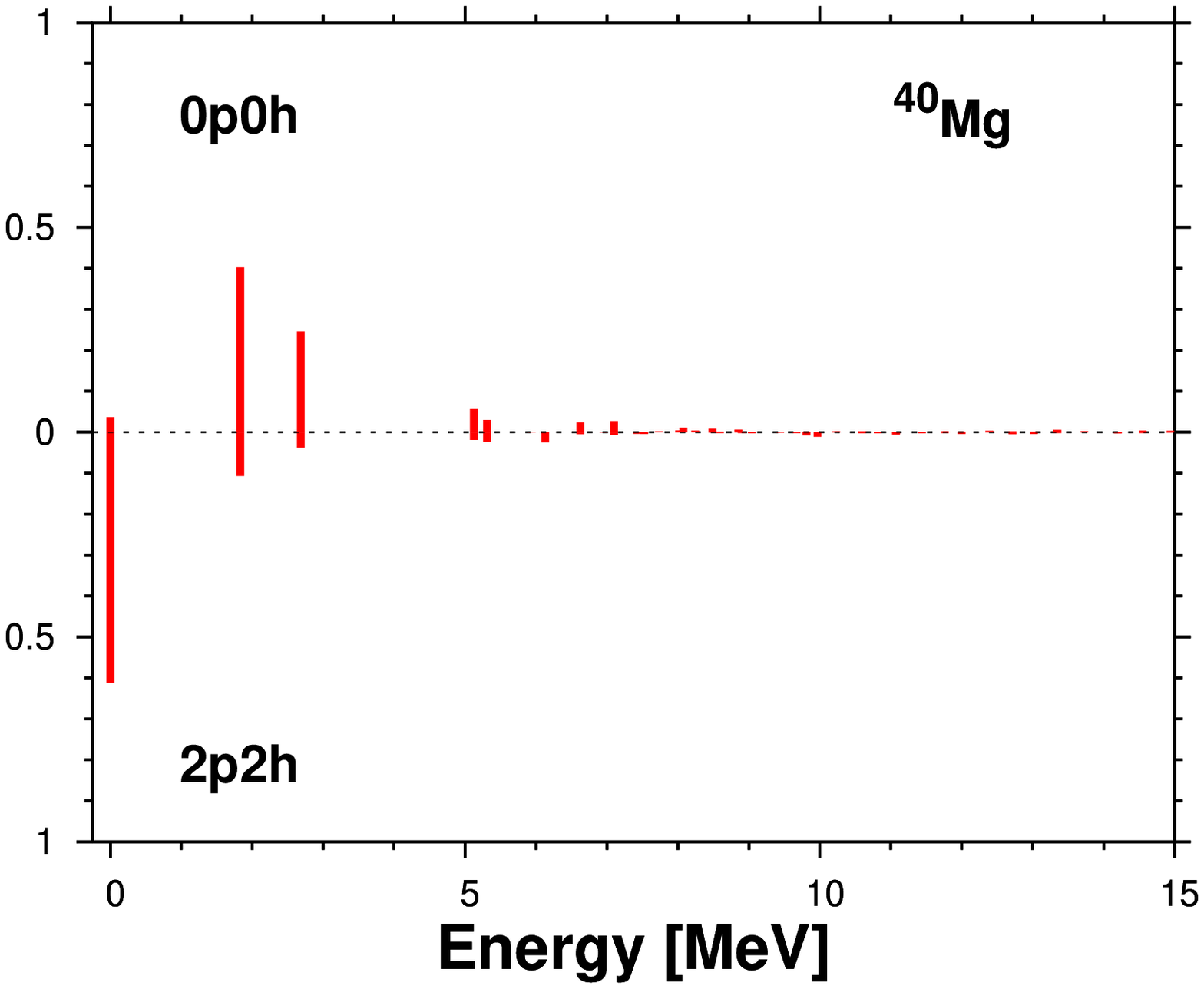} 
    \caption{Overlaps (squared) of the closed shell 0$^+$ (upper panels) and the intruder 
     0$^+$ (bottom panels) with the physical 0$^+$ states in $^{46}$Ar, $^{44}$S, $^{42}$Si, and
    $^{40}$Mg}
    \label{fig:over2}
\end{center}
\end{figure}

What happens to these bands (or states) when full mixing is allowed?
Or, in other words, what is the structure of the physical spectra
according to our calculations? This aspect is better explained by
figures~\ref{fig:over1} and \ref{fig:over2}, where we have plotted the
projection of the the lowest closed shell 0$^+$ state (upper panel) and the
lowest intruder 0$^+$ state (bottom panel) onto the physical  0$^+$ states. The
projections are obtained by means of the Lanczos strength function
method.  The spikes give the squared overlap of these two states with
the 
physical states.  In figure \ref{fig:over1} we present the results for
$^{50}$Ti $^{52}$Cr and $^{54}$Fe and in figure~ \ref{fig:over2} for
$^{46}$Ar, $^{44}$S, $^{42}$Si, and $^{40}$Mg.

The results for all the isotopes with Z$>$20 are quite similar; The
closed shell configuration is dominant in the ground state (about
70\%), nothing unexpected. What may be a bit more surprising is that
the intruder 0$^+$ state dominates the first excited 0$^+$ state, even
more strongly (about 75\%). The same happens for the other spin
values. The lowest physical state of each spin is dominated by the closed shell
configuration while the intruder state has large overlap with  a single
physical state. This may  indicate  that  these intruder structures
can survive to the mixing with the huge number of configurations in
the model space, keeping their identity as collective yrare bands, and
providing a good example of coexistence of spherical and deformed
shapes.

In Figure~\ref{fig:over2}, $^{46}$Ar shows already a larger degree of mixing 
between the closed shell and the 2p-2h bandhead, nonetheless, the
former is  still dominant in the physical ground state. The 2p-2h
0$^+$, is split between the 0$^+_2$ and 0$^+_3$ physical states.
The ground state of  $^{44}$S is an even mixture of the (spherical)
closed shell 0$^+$, and the deformed 2p-2h  0$^+$ bandhead. The low lying
0$^+_2$ has the same amount of mixing. This is a  rather unusual
situation provoked by the near degeneracy of
the two states before mixing. A similar situation occurs in  $^{42}$Si,
except that now the 2p-2h 0$^+$ does not correspond to a deformed
shape. On the contrary, the ground state of  $^{40}$Mg is fully
dominated by the  deformed 2p-2h  0$^+$ bandhead. This dominance is
maintained up to J=6. Hence a very simple picture emerges; the
deformed intruder band that was yrare near the stability becomes yrast
at the neutron drip line; the shape transition has finally
happened. This kind of inversion of spherical  and deformed
configurations, leading to shape transitions,
takes also place in the other neutron rich semi-magic Magnesium isotope
 $^{32}$Mg (N=20).

 Another striking prediction of our calculations is the occurence of a
 superdeformed band in $^{42}$Si of 4p-4h nature. Contrary to the
 2p-2h structure analysed above, that wasn't deformed due to the
 1d$_{5/2}$ proton orbit closure, the 4p4h configuration has open
 shell protons and neutrons and attains a large deformation.  In
 addition, when mixing in the full space is allowed, the band stands
 with even enhanced deformation.  The rotational bandhead is located
 at an excitation energy of 5 MeV.  The band is caracterized by an
 intrinsic quadrupole moment $Q_0$ of 86~$e~fm^2$ (corresponding to a
 deformation parameter $\beta \sim 0.5$) and terminates at spin
 8$\hbar$.

Finally, we gather  in Table~\ref{tab:n28} some numerical
results for the Z$<$20 isotones. Notice in particular the near degeneracy of the first 
excited 2$^+$ and  0$^+$ states in $^{44}$S and  $^{42}$Si and the complete disappearance of
the closed shell component in the ground state of $^{40}$Mg already
apparent in Figure~\ref{fig:over2}.

\begin{table}
\caption{N=28 isotones: Calculated spectra, quadrupole properties and occupancies
 of the Z$<$20, N=28, isotones.}
\label{tab:n28}       
 \begin{tabular*}{\linewidth}{@{\extracolsep{\fill}}lcccc} 
\hline\noalign{\smallskip}
& $^{40}$Mg 
& $^{42}$Si 
& $^{44}$S
& $^{46}$Ar  \\
\noalign{\smallskip}\hline\noalign{\smallskip}
E$^*$(2$^+$)(MeV)   & 0.81   & 1.49     & 1.22 &  1.51     \\
E$^*$(4$^+$)   & 2.17   & 2.68     & 2.25 &  3.46      \\
E$^*$(0$^+_2$) & 1.83   & 1.57     & 1.26 &  2.93    \\
$Q(2^+$)(e~fm$^2$)   & -21   & 16      & -17    & 20  \\
B(E2)(e$^2$~fm$^4$) & 108    & 71     & 93   & 93    \\
$\langle {n}_{7/2}\rangle$  & 5.54 & 6.16 &  6.16  & 6.91  \\
$({f}_{7/2})^8$ \% & 3  & 28   & 24   & 45  \\
\noalign{\smallskip}\hline
 \end{tabular*}
\end{table}

\section{Collective yrare bands in $^{52}$Cr and $^{54}$Fe}

The occurrence of yrare bands of enhanced collectivity, with a well
defined particle-hole structure on top of the ground state, has been
recently documented by experiments in $^{56}$Ni \cite{rudolf},
$^{36}$Ar \cite{carlsv} and $^{40}$Ca \cite{idegu}. In the Nickel and
Argon cases the bands are dominantly made of 4p-4h excitations while
in $^{40}$Ca the leading structure is 8p-8h \cite{ca40sd}. In
$^{56}$Ni the band is classified as highly deformed $\beta$= 0.3/0.4 ,
while in $^{36}$Ar and $^{40}$Ca they are characterised as
superdeformed with $\beta$= 0.4/0.5 and $\beta$= 0.5/0.6 respectively.
In a recent study \cite{mizu}, Mizusaki and co-workers have carried
out a theoretical search for this kind of bands in the N=28 isotopes,
using the Monte Carlo shell model approach. According to their
calculations with the FPD6 interaction~\cite{Richter.ea:1991} the
2p-2h collective bands of $^{52}$Cr and $^{54}$Fe, that we have
studied in the preceding section persist in the full calculation
(actually it is not very clear in their paper if the results are from
MCSM calculations or from direct diagonalizations, that 
should involve truncations at least in $^{54}$Fe). They
discuss also the possible existence of another yrare band in
$^{54}$Fe, this time of 4p-4h nature. Their conclusion is positive,
but based only on a mean field calculation with incorporates variation
after projection (VAP) because (sic) the MCSM is impractical in this
case. We will reexamine this problem as a natural continuation of our
present study.

 The calculation of yrare bands in nuclei with  large shell model
 dimensionalities, poses serious computational problems. The reason
 is that, most often, the interesting states lie at  excitation
 energies  where the level density is high. This represents a real
 challenge for all the shell model approaches, because it requires the
 calculation of many states of the same spin and parity. While this
 can be easily done for m-scheme dimensions of  a few tens of millions, the
 task becomes formidable for dimensions of hundreds of millions. The
 method of  ref~\cite{mizu} relies in the existence of local minima in
 the projected energy surfaces resulting of a constrained Hartree-Fock
 calculation. The Slater determinants corresponding to these minima
 are then fed into the MCSM.

 In our approach, we first compute as many fully converged 0$^+$
 states as necessary to obtain the band-heads of the potential yrare bands. To
 select them, we apply the quadrupole operator and use the LSF method
 to decide if there is a dominant branch connecting them to a 2$^+$
 state that would then be a member of the band. We proceed upwards in
 spin until the collective transition strength diminish drastically or
 bifurcates.  In the case of the 2p-2h bands we have already shown  that 
 the band-heads are the first excited 0$^+$'s.
 
 In $^{52}$Cr the calculations are carried out in the full $pf$-shell
 (m-scheme dimension 45 millions). The 2p-2h band-head at 2.43~MeV, is
 in correspondence with the experimental 0$^+_2$ at 2.65~MeV. It has a
 46\% content of 2p-2h, 26\% of 3p-3h, 18\% of 4p-4h and minor
 percentages of the other components.  Besides the ``coherent'' 2p-2h
 state, the next four 0$^+$ states produced by the calculation are
 also in correspondence with the experimental 0$^+$'s:
\begin{center} 
 0$^+_3$; exp. 4.74~MeV; th. 5.15~MeV,

 0$^+_4$; exp. 5.60~MeV; th. 5.47~MeV,

 0$^+_5$; exp. 5.75~MeV; th. 6.04~MeV,

 0$^+_6$; exp. 6.10~MeV; th. 6.30~MeV,
\end{center}
 providing a rather spectacular showpiece  of  shell model spectroscopy.

\begin{table}[b]
\begin{center}
\caption{Yrare 2p\,-2h  band in $^{52}$Cr. Excitation energies are
  relative to the 0$^+_2$ band-head, experimentally at 2.64~MeV and
  calculated at 2.43~MeV. Energies in MeV,  B(E2)'s in $e^2\,fm^4$,
  Q's in $e\,fm^2$}
\label{tab:cr52yrare}
\vspace{0.3cm}
\begin{tabular*}{\textwidth}{@{\extracolsep{\fill}}cccccc}
\hline\hline
J & $\Delta$E & \% of B(E2)$\uparrow$ & \% of 2p\,-2h &
B(E2)$\downarrow$ & Q$_{spec}$ \\
\hline
0 & 0.00 &    & 46 &     &      \\
2 & 0.44 & 77 & 43 & 238 & -27.7\\
4 & 1.32 & 82 & 43 & 323 & -39.3\\
6 & 2.51 & 77 & 42 & 329 & -39.3\\
8 & 4.14 & 83 & 44 & 331 & -39.2\\
10& 6.03 & 66 & 39 & 239 & -25.5\\
12& 8.02 & 27 & 51 &  66 & -13.7\\
12& 8.31 & 28 &    &  69 &      \\
\hline\hline
\end{tabular*}
\end{center}
\end{table}

  Let's see how the procedure works. First we apply 
  the quadrupole operator to the   0$^+$ bandhead: $Q | 0^+ \rangle$. This
  operation generates a
  threshold vector that we take as the ``pivot'' in the Lanczos
  procedure. We perform N$\sim$100 iterations to produce 
  an approximate strength
  function. At this stage, either there is one state that carries most
  of the strength, thus belonging to the band, or not, in
  which case there is no band. In the positive case the resulting
   $|2^+\rangle$
  state is retained as a band member and we proceed to act  with $Q$
  on it. In this  step, there are  several
  possible angular momentum couplings,
  and  we follow the  $\Delta$J=2 path in the
  even-even nuclei. The procedure is repeated until the strength
  bifurcates, dilutes or plainly disappears.

  In  $^{52}$Cr we can definitely speak of a ``theoretical'' yrare
  band. Our results for the excitation energies are very close to
  those of ref.~\cite{mizu}. In
  table \ref{tab:cr52yrare} we list  the
  excitation energies, the percentage of the B(E2)$\uparrow$ of the
  transition from J-2 to J carried by each state, the amount of 2p-2h
  components, the B(E2)$\downarrow$ and the spectroscopic quadrupole
  moments. All these numbers are consistent with a deformed band up to
  J=10, afterwards a bifurcation and reduction of the strength takes
  place. Notice the large B(E2) values, similar to those of the yrast
  band of $^{48}$Cr, corresponding to $\beta \sim$~0.3. Besides, the
  energy spacings are very close to those of a rigid rotor;
 
\medskip
\begin{center}
 $\displaystyle{\frac{\Delta E(4^+)}{\Delta E(2^+)}}$=3(3.33);

 $\displaystyle{\frac{\Delta E(6^+)}{\Delta E(4^+)}}$=1.9(2.1);  

 $\displaystyle{\frac{\Delta E(8^+)}{\Delta E(6^+)}}$=1.65(1.71);  

 $\displaystyle{\frac{\Delta E(10^+)}{\Delta E(8^+)}}$=1.46(1.53);  
\end{center}
\medskip

\noindent
 (the values in parenthesis are those of the rigid rotor limit). The
 percentage of the 2p-2h components as well as the occupation
 numbers of the individual orbits are nicely constant too. The
 spectroscopic quadrupole moments correspond to 
 a prolate rotor and are fully consistent with the B(E2)'s.
 The sequence of calculated states, 0$^+$ at 2.43~MeV, 2$^+$ at
 2.87~MeV, 4$^+$ at 3.75~MeV and 6$^+$ at 4.94~MeV can be put in
 correspondence with the experimental ones at  2.64~MeV (0$^+$), 
 3.16~MeV (2$^+$),  4.04~MeV (4$^+$) and  5.14~MeV (6$^+$). However,
 there are no transitions experimentally observed linking these
 states. This means that, due to the presence of many other 2p-2h
 states to which to decay by M1 transitions and to the phase space
 enhancement of the E2 transitions to the yrast band, the
 ``theoretical'' yrare band does not show up experimentally as
 a $\gamma$-cascade.

 We have also searched for other bands of 4p-4h nature but we have
 found none compatible with our requirements.

\begin{table}[htb]
\begin{center}
\caption{Yrare 2p\,-2h  band in $^{54}$Fe. Excitation energies are
  relative to the 0$^+_2$ band-head, experimentally at 2.56~MeV and
  calculated at 2.77~MeV. Energies in MeV,  B(E2)'s in $e^2\,fm^4$,
  Q's in $e\,fm^2$}
\label{tab:fe54yrare}
\vspace{0.3cm}
\begin{tabular*}{\textwidth}{@{\extracolsep{\fill}}ccccc}
\hline\hline
J & $\Delta$E & \% of 2p\,-2h &
B(E2)$\downarrow$ & Q$_{spec}$ \\
\hline
0 & 0.00 &  45 &     &      \\
2 & 0.55 &  43 & 232 & -29.1\\
4 & 1.57 &  37 & 262 & -29.9\\
6 & 3.07 &  40 & 298 & -40.5\\
8 & 4.00 &  35 & 205 & -22.4\\
\hline\hline
\end{tabular*}
\end{center}
\end{table}

 For $^{54}$Fe the full space calculations of the different strength
 functions demand a computational effort beyond our present means. However, we have verified in $^{52}$Cr
 and in other cases that it is accurate enough to compute them at the
 truncation level t=n+4 (t is the number of particles that can be
 excited from the 1f$_{7/2}$ orbit to the rest of the $pf$-shell and n
 is the np-nh character of the band). Hence, we can use t=6 for the 2p-2h
 and t=8 for the 4p-4h cases (m-scheme
 dimension 177 millions).  The situation for the lowest yrare band
 resembles that of $^{52}$Cr; the coherent 2p-2h state represents 43\%
  of the 0$^+_2$ at 2.77~MeV (exp. at 2.56~MeV).
  The yrast band is also well reproduced by the calculation with the KB3G interaction;
  the states 2$^+$ at
  1.52~MeV, 4$^+$ at 2.26~MeV and 6$^+$ at 3.02~MeV agree nicely with
  the experimental ones at   
  1.41~MeV (2$^+$),  2.54~MeV (4$^+$) and  2.95~MeV (6$^+$).
  The results for the yrare band are gathered in
  Table~\ref{tab:fe54yrare}. 
  The 2$^+$-0$^+$ splitting of 0.55~MeV can be put in correspondence
  with the experimental 2$^+_3$-0$^+_2$ splitting (0.61~MeV). As in the
  case of $^{52}$Cr, there is no doubt about the collective character
  of this ``theoretical'' yrare band, even is some rotational features
  are less perfect than in $^{52}$Cr, for instance, the band is
 completely disolved beyond J=8. No in-band transitions are
  experimentally seen, for the same reasons advocated for $^{52}$Cr.

  We have made a very thorough  search for more deformed yrare bands of
  4p-4h character also in $^{54}$Fe, motivated by the prediction made
  in ref~\cite{mizu} using the FPD6 interaction and a deformed
  Hartree-Fock calculation with variation after projection to good
  angular momentum (VAP). Indeed, if we do a calculation with the same
  interaction and a fixed 
  number of particles outside the 1f$_{7/2}$ orbit equal to four, we find a
  collective band with a B(E2)(2$^+$
  $\rightarrow$ 0$^+$)$\approx$ 400~e$^2$~fm$^4$ which is not far from
  the VAP result  $\approx$ 500~e$^2$~fm$^4$. However, when  we perform a
  --nearly- full calculation (t=8), fully converging all the 0$^+$
  and 2$^+$ states up to 7~MeV of excitation energy (eight 0$^+$'s and
  fifteen 2$^+$'s)
  the 4p-4h band members are completely fragmented among the physical
  states  and no trace of any other collective structure is found in the
  results. With the only caveat of the non-completeness of  the calculation 
  in the $pf$-shell,
  we can refute  the VAP prediction, that, obviously, is far from
  including  the mixing present in the shell model calculations.

  In summary, we have studied the occurrence of collective  yrare
  bands of well defined  p-h
  character in the N=28 isotones. We have confirmed the existence of
  deformed 2p-2h  yrare bands  in  $^{52}$Cr and  $^{54}$Fe. However
  they  don't
  show up experimentally as such;  
  the states decay preferentially out of the band due to the phase
  space enhancement. In  $^{40}$Mg the
  2p-2h band actually becomes yrast, producing a spherical to deformed
  shape transition at the drip line, while the N=28 closure vanishes.

{\bf Acknowledgements}.
 This work has been partly supported by MCyT~(Spain), grant BFM2000-30
 and by the IN2P3~(France)-CICyT~(Spain)
 agreements. We also thank the CCC-UAM for a computational grant.

\end{document}